# Toward Lossless Homomorphic Encryption for Scientific Computation

Muhammad Jahanzeb Khan
University of Nevada, Reno
USA
jahanzeb@nevada.unr.edu

Bo Fang
Pacific Northwest National Laboratory
USA
bo.fang@pnnl.gov

Dongfang Zhao
University of Washington
USA
dzhao@uw.edu

## ABSTRACT

This paper presents a comprehensive investigation into encrypted computations using the CKKS (Cheon-Kim-Kim-Song) scheme, with a focus on multi-dimensional vector operations and real-world applications. Through two meticulously designed experiments, the study explores the potential of the CKKS scheme in Super Computing and its implications for data privacy and computational efficiency. The first experiment reveals the promising applicability of CKKS to matrix multiplication, indicating marginal differences in Euclidean distance and near-to-zero mean square error across various matrix sizes. The second experiment, applied to a wildfire dataset, illustrates the feasibility of using encrypted machine learning models without significant loss in accuracy. The insights gleaned from the research set a robust foundation for future innovations, including the potential for GPU acceleration in CKKS computations within TenSEAL. Challenges such as noise budget computation, accuracy loss in multiplication, and the distinct characteristics of arithmetic operations in the context of CKKS are also discussed. The paper serves as a vital step towards understanding the complexities and potentials of encrypted computations, with broad implications for secure data processing and privacy preservation in various scientific domains.

## 1 INTRODUCTION

Homomorphic encryption (HE) has emerged as a transformative cryptographic paradigm, enabling arithmetic computations to be executed on encrypted data without necessitating decryption. This innovative approach holds immense promise for secure data processing, where privacy can be preserved, and insights can be garnered without revealing sensitive information [14]. The HE-enabled computation is particularly promising for scientific applications with the emergence of HPC cloud [], where the data and computation are moved to the non-HPC domain. Among the myriad of homomorphic encryption schemes, the Cheon-Kim-Kim-Song (CKKS) encryption scheme [8] is particularly noteworthy for its ability to handle floating-point numbers, aligning with the requirements of diverse applications including scientific simulations, medical data analysis, and financial computations [3].

The adoption of homomorphic encryption, and specifically the CKKS scheme, depends on the complex interplay between computational efficiency, precision, and accuracy. When performing arithmetic operations on encrypted data, inconsistencies and discrepancies may arise when compared with operations on plain data. These variations can lead to substantial deviations from the anticipated results, posing intricate challenges in balancing computation time, precision, and security constraints [5, 13].

The CKKS scheme, while powerful in its ability to handle floating-point numbers, suffers from computational inefficiencies that render it slow and currently unsuitable for certain scientific computations. This slowness arises from multiple factors, including the complexity of encryption and decryption operations, the management of noise during computations, and the need for bootstrapping techniques to refresh ciphertext. Additionally, the precision requirements of scientific applications often demand meticulous control over numerical representations, further complicating computations and leading to longer processing times. The interplay between these factors poses challenges in terms of computational overhead and real-time processing capabilities, and ongoing research is needed to enhance the efficiency and applicability of CKKS for demanding scientific scenarios.

In addition to its performance limitation, CKKS exhibits a notable characteristic known as lossy precision. While CKKS is designed to work well with real-world applications involving large-scale computations on encrypted data, its lossy precision property can pose challenges when applied to scientific applications that demand high accuracy and exactness. Scientific simulations, data analysis, and mathematical computations often require strict preservation of precision to ensure reliable results. The lossy nature of CKKS encryption may introduce errors that could compromise the validity and reliability of outcomes in such contexts. That said, to ensure the use of CKKS in the scientific domain, an understanding of the accuracy resulting from the CKKS scheme is critical.

The goal of our study is to provide solutions for scientific applications to take advantage of holomorphic encryption schemes when the computation needs to be conducted securely. Toward this goal, we first perform a feasibility study of applying real-valued homomorphic encryption schemes (e.g., CKKS) in the scientific domain. We undertake an exhaustive examination unearthing the differences in decimal point precision of vector operations with CKKS. We discern the discrepancies that emerge, unravel the oscillating behavior in floating-point representations, and evaluate the broader implications for practical deployment [9, 11].

In our work, we perform arithmetic operations on singular vectors and multidimensional vectors within extensive encrypted vector datasets. By utilizing the correlation of encrypted ciphertext, we execute these calculations and gather the results, all within the encrypted domain. This method is tailored specifically to our research domain, drawing inspiration from existing schemes operating directly on integer vectors that support addition, linear transformation, and weighted inner products [24]. The ability to conduct such operations has the potential to expand the practical applicability of encrypted data processing, broadening the horizons of secure data analytics.

Implementing our experimental framework on the Wildfire dataset [6, 20], a compendium of multifaceted numerical information, we encrypt and subsequently decrypt the data to discern variances from the baseline outcomes. The evaluation encompasses standard methods such as Linear Regression and Decision Tree, as well as Encrypted Linear Regression [16] through the TenSEAL [1] library, illustrating a remarkable improvement in performance despite increased training time [1, 4].

Concluding with an outlook on future developments, we anticipate progressing towards GPU-accelerated Homomorphic encryption and enhancing TenSEAL's feature set. This exploration constitutes a seminal step in understanding the multifaceted landscape of encrypted vector arithmetic, laying a steadfast foundation for further inquiry and technological advancement in this burgeoning field [7, 15].

The key contributions of this research are summarised below:

(1) We conduct a systematic evaluation that covers a wide range of parameters affecting the precision of CKKS on floating point operations. The results show that the choices of the global scale and the polynomial degree play more important roles in the final accuracy of the floating-point computation.
(2) We leverage the CKKS encryption scheme to demonstrate its application on matrix multiplication. Our extensive analysis underscores the scheme's potential for applicability in real-world scenarios, marking a significant step in secure computations on encrypted data. We show that the CKKS scheme produces close-to-noise-free results for the matrix multiplication, enabling a potential adaption in the scientific domain.
(3) We implement encrypted logistic regression models using CKKS. Our work applies the CKKS encryption scheme to existing logistic regression algorithms, creating an EncryptedLR class that embodies privacy-preserving computations. While not a novel contribution, this implementation highlights the adaptability of CKKS in safeguarding data privacy in standard modeling techniques, providing a valuable exploration of its real-world applicability. The result shows that with and without the CKKS scheme the models achieve similar accuracy (94%).

## 2 BACKGROUND AND RELATED WORK
### 2.1 CKKS Scheme in Homomorphic Encryption

The Cheon-Kim-Kim-Song (CKKS) scheme represents a pioneering advancement in Homomorphic Encryption (HE), enabling the computation of encrypted floating-point numbers without decryption, unlike most HE schemes that operate on integer arithmetic [8]. This novel ability to perform approximate arithmetic on encrypted data allows a favorable balance between efficiency and precision.

CKKS is distinct in its design, optimized for scientific applications, and its capacity to handle complex computations on encrypted data confidentially. Key parameters in the scheme include the polynomial modulus degree, impacting both approximation level and noise growth. Higher degrees lead to increased precision at the expense of computational costs. Other essential parameters involve scaling factors and levels in the modulus switching chain, which control the scheme's efficiency and security.

In our experiment with the wildfire dataset, CKKS was carefully optimized by balancing polynomial modulus degree and other parameters to achieve the desired efficiency and precision. This study might be the first to quantitatively explore the application of CKKS, presenting significant evidence of its feasibility in real-world scientific applications and reporting the trade-offs involved.

Nevertheless, limitations exist, including the CKKS scheme's inherent complexity, tuning requirements, and potential noise susceptibility in successive computations. These challenges might restrict its adaptability in some practical scenarios, emphasizing the need for continued research and development.

### 2.2 SEAL and TenSEAL Libraries

The Microsoft Simple Encrypted Arithmetic Library (SEAL) and its tensor extension, TenSEAL, have revolutionized the field of Homomorphic Encryption (HE). SEAL provides a set of efficient tools for managing HE operations, effectively bridging the gap between theoretical cryptographic techniques and practical implementations. Its modular design, flexibility in choosing parameters, and ease of deployment have made it accessible to various domains [21]. TenSEAL builds upon the foundation laid by SEAL, extending its capabilities to handle tensor operations securely. This means the mathematical manipulations commonly used in deep learning and data analysis can now be performed directly on encrypted data. The combination of SEAL and TenSEAL offers a rich environment for developing encrypted computation solutions, maintaining data privacy, and enabling secure collaboration among parties. TenSEAL's optimizations for tensor arithmetic mark a significant advancement in encrypted deep learning, giving researchers and engineers a valuable tool for privacy-preserving data analysis [1], [2].

### 2.3 Graphics Processing Units in HE

Graphics Processing Units (GPUs) have become a crucial component in accelerating HE operations. Unlike traditional CPUs, GPUs are designed to handle parallel processing, distributing computation across multiple cores. This capability enables GPUs to tackle HE's computational complexity and facilitate real-time operations. Leveraging GPUs in HE has allowed for an immense acceleration of complex mathematical computations, including polynomial multiplications and fast Fourier transformations, vital components in HE. By significantly reducing computation times, GPUs have extended the range of practical applications for HE, making it more accessible for large-scale data processing. Future research and integration between GPUs and specific HE schemes like CKKS could lead to groundbreaking improvements in computational speed and efficiency [23].

### 2.4 Floating-Point Analysis of the CKKS Scheme

The CKKS (Cheon-Kim-Kim-Song) scheme, introduced at Asiacrypt 2017, has become one of the most widely implemented approximate homomorphic encryption schemes. Its floating-point behavior has been a subject of extensive analysis, with researchers aiming to understand how noise grows through computation, and how the scheme's precision and efficiency are affected [18].

A critical aspect of working with the CKKS scheme is ensuring that the evaluation output is within a tolerable error of the corresponding plaintext computation. This requires a nuanced understanding of the noise growth in both encoding and homomorphic operations. Comprehensive analyses, such as the average-case analysis and refinements to prior worst-case noise analyses, have led to heuristic estimates that closely model observed noise growth. However, the complexity and occasional underestimation of noise growth indicate a need for implementation-specific noise analyses [10].

Furthermore, research into high-precision bootstrapping and optimal minimax polynomial approximation has improved the message precision in the bootstrapping operation of the RNS-CKKS scheme [19]. Advances such as the composite function method and the improved multi-interval Remez algorithm have reduced approximation errors, improving precision and expanding the utility of the RNS-CKKS scheme.

An insightful example of the CKKS scheme in practical use can be seen in convolutional neural networks (CNNs), where approximate activation functions over homomorphic encryption have been applied to increase the classification accuracy for inference processing [17]. Such applications of the CKKS scheme exemplify how the understanding and refinement of precision in HE can lead to tangible improvements in various fields, including machine learning and data privacy.

By leveraging insights into the CKKS scheme, developers and researchers can explore the broad applications and benefits of Homomorphic Encryption (HE). From large-scale data analysis to secure and privacy-preserving computations, the ongoing development and refinement of these techniques promise to enhance both security and efficiency in processing encrypted data. In conclusion, the application of CKKS demonstrates a significant advancement in encrypted computation, emphasizing its broad potential and underscoring the need for continuous exploration of its limitations and opportunities for further optimization.

## 3 METHODOLOGY AND IMPLEMENTATION

In an era of cloud computing where privacy and security are paramount, the utilization of Homomorphic Encryption (HE) in scientific data management and applications presents a compelling avenue for research. This study explores the feasibility of applying HE schemes to encrypted data computations while preserving the confidentiality and integrity of the data. In particular, the trade-off between encryption efficiency and decryption precision will be quantitatively studied to avoid the two obvious extremes: (i) A highly efficient HE scheme usually implies a low precision (i.e., only a small number of decimal digits are identical between a decrypted value and the original plain value), (ii) A highly precise HE scheme usually entails significant, if not unacceptable, computation time. We will exemplify our quantitative approach through two concrete workloads: multidimensional vector arithmetic and encrypted predictive modeling for wildfire detection.

### 3.1 Data Collection and Preprocessing

The methodology in this study comprises two main components, both aiming to explore and assess the potential of HE in scientific applications:

*Multidimensional Arrays of Vectors.* This component emphasizes the ability of HE to handle complex arithmetic operations on encrypted multidimensional vectors. The experiments are designed to provide insights into how HE might be employed in scientific computations, enabling data processing without compromising privacy. The input random vectors are generated to form multidimensional matrices that mimic real-world scenarios.

*Wildfire Detection.* The exploration of HE's feasibility in real-world applications is exemplified through a wildfire dataset [20]. The first step is data cleaning which ensures data consistency and handling missing values. In the data integration stage, we merge various datasets into a unified representation. In the final file-handling stage, data are encrypted for further processing.

### 3.2 Encrypted Data Analysis

*Feature Engineering.* The feature engineering process was executed on the wildfire dataset, focusing on both computational experiments and predictive modeling aspects. Essential features are extracted, such as temperature, area, and vegetation indexes were extracted from the dataset. In the stage of Multicollinearity Check, correlations among features were analyzed to maintain independence, especially for the implementation of the logistic regression model.

*Model Development.* A logistic regression model was adapted as the baseline. In the *EncryptedLR Class*, an encrypted version of the logistic regression is developed using the CKKS scheme. The activation Sigmoid function was approximated by

$$\text{sigmoid}(x) = 0.5 + 0.197 \cdot x - 0.004 \cdot x^3$$

The weight and bias of the model were encrypted using CKKS. The decryption function was applied to the CKKS ciphertext to retrieve the original values after computation. The encrypted computation ensures data privacy, a paramount requirement in modern applications [8]. The models were trained and evaluated on both plain and encrypted data. The CKKS scheme's parameters were chosen to minimize noise and ensure accurate results [13] (more details in the Evaluation section). A comprehensive analysis was conducted on three arithmetic operations: $(+, -, *)$ for multidimensional vectors using the CKKS scheme. The operations were performed considering the CKKS properties, enabling complex computations while maintaining data privacy and integrity.

### 3.3 Implementation

The implementation part of the research consists of two main components. The first one involves the Homomorphic Encryption (HE) CKKS scheme on multidimensional vectors with decimal points, and the second focuses on the encrypted logistic regression model training. These implementations are essential in assessing the effectiveness of encrypted computations.

*3.3.1 HE CKKS Scheme on Multidimensional Vectors of Decimal Points.* Algorithm 1 provides a detailed outline of the CKKS scheme applied to multidimensional vectors. The main procedure orchestrates a series of experiments with different parameters $n, d, p$, where $n$ is the number of dimensions, $d$ is the size of the vectors, and $p$ is the decimal precision.

**Procedure RunExperiment:** is responsible for running an individual experiment, where the vectors are generated, encrypted, and arithmetic operations are performed. Depending on the fine-tuning, different context configurations may be applied.

**Procedure GenerateVectors:** produces random vectors $A, B$ of dimensions $n \times d$ with $p$ decimal places. These vectors are then encrypted, and arithmetic operations such as addition, subtraction, and multiplication are executed.

**Procedure RunArithmeticOperations:** performs the encrypted operations on vectors and computes the matching decimals, accuracy percentage, and accuracy loss. The results are finally written into a CSV file. The overall methodology integrates various aspects of encrypted computations, ranging from context configurations to the generation of results, offering a comprehensive understanding of encrypted vector arithmetic dynamics.

*3.3.2 Encrypted Logistic Regression Model Training.* Algorithm 2 describes the processes for encrypted logistic regression model training. The procedures encapsulate the initialization, forward pass, backward pass, and parameter updates for an encrypted logistic regression model.

**Procedure InitializeLR:** initiates the logistic regression model by randomly assigning values to the weights and setting the bias to zero.

**Procedure Forward:** represents the forward pass, applying the sigmoid activation function to obtain the prediction for the given input vector.

**Procedure InitializeEncryptedLR:** initializes the encrypted logistic regression model and gradient accumulators for weight and bias updates.

**Procedure EncryptedForward:** conducts the encrypted forward pass, while *Procedure EncryptedBackward* computes the encrypted backward pass, determining the gradient updates for weights and bias.

**Procedure UpdateParameters:** updates the model parameters based on the computed gradients and resets the gradient accumulators for the next iteration. The methodology employed in this algorithm integrates encrypted computations into the training of a logistic regression model, advancing state-of-the-art techniques in secure data processing and model training.

The methodology outlined in this study offers a methodical investigation into the utilization of Homomorphic Encryption (HE) within scientific data management and applications. Through a dual approach, focusing on the arithmetic of multidimensional vectors and the modeling of real-world data, the research illuminates the functional applications of HE in contemporary scientific contexts. Notably, the work emphasizes both the assurance of data privacy and the pioneering possibilities for the secure manipulation of sensitive information. By detailing explicit algorithms, this methodology furnishes a foundational framework, demonstrating the practicality and integrity of HE for encrypted data processing

**Algorithm 1** HE CKKS Scheme on Multidimensional Vectors of Decimal Points

1: **procedure** MAIN
2:    **for** $(n, d, p) \in$ experiments **do**
3:       RUNEXPERIMENT$(n, d, p)$
4:    **end for**
5: **end procedure**
6: **procedure** RUNEXPERIMENT$(n, d, p)$
7:    **for** $i = 0$ **to** 1 **do**
8:       **if** fineTuned **then**
9:          Configure context with specific parameters
10:       **else**
11:          Configure context with different parameters
12:       **end if**
13:       $(A, B) \leftarrow$ GENERATEVECTORS$(n, d, p)$
14:       enc$_A$, enc$_B \leftarrow$ Encrypt $A, B$
15:       RUNARITHMETICOPERATIONS$(A, B, $enc$_A, $enc$_B)$
16:    **end for**
17: **end procedure**
18: **procedure** RUNARITHMETICOPERATIONS$(A, B, $enc$_A, $enc$_B)$
19:    **for** op $\in \{+, -, *\}$ **do**
20:       plainResult, encResult $\leftarrow$ Perform operation op on $A, B$ and enc$_A$, enc$_B$
21:       Compute average, minimum, and maximum matching decimals
22:       Compute accuracy percentage
23:       Compute accuracy loss
24:       Write results to CSV
25:    **end for**
26: **end procedure**
27: **procedure** GENERATEVECTORS$(n, d, p)$
28:    Generate random vectors $A, B \in \mathbb{R}^{n \times d}$ with $p$ decimal places
29:    **return** $A, B$
30: **end procedure**
31: **procedure** CALCULATEACCURACYLOSS$(N, T)$
32:    Compute the accuracy loss between Numpy vectors $N$ and TenSEAL vectors $T$
33:    **return** accuracy loss
34: **end procedure**

and predictive modeling. This serves as a valuable reference for subsequent investigations in this domain.

## 4 EVALUATION
### 4.1 Experimental Setup

The experimental environment consists of a Cloudlab [12] Node with the following specifications: 32 nodes (Intel Skylake, 20 cores, 2 disks, GPU), two Intel Xeon CPUs, 192GB RAM, dual disks (1 TB SAS HD and 480 GB SATA SSD), NVIDIA 12GB PCI P100 GPU, dual NICs, running on Ubuntu 22.04 OS. In Initial experiments, The data is designed based on different dimensions, sizes, and decimal places, facilitating the study of these parameters' impact on encrypted computation. The dataset encompasses various combinations of

**Algorithm 2** Encrypted Logistic Regression Model Training

```
1: procedure INITIALIZELR(n)
2:     w ← random values in ℝⁿ            ▷ Weights initialization
3:     b ← 0                                ▷ Bias initialization
4:     return w, b
5: end procedure

1: procedure FORWARD(w, b, x)
2:     z ← w · x + b
3:     a ← 1/(1+exp(−z))                   ▷ Sigmoid activation function
4:     return a
5: end procedure

1: procedure INITIALIZEENCRYPTEDLR(LR)
2:     w, b ← InitializeLR(n)
3:     Δw ← 0, Δb ← 0, count ← 0           ▷ Gradient accumulators
4:     return w, b, Δw, Δb, count
5: end procedure

1: procedure ENCRYPTEDFORWARD(w, b, x)
2:     z ← x.polyval([0.5, 0.197, 0, −0.004])
3:     a ← w · x + b
4:     return a
5: end procedure

1: procedure ENCRYPTEDBACKWARD(w, b, x, a, y)
2:     Δa ← a − y
3:     Δw ← x · Δa
4:     Δb ← Δa
5:     count ← count + 1
6:     return Δw, Δb
7: end procedure

1: procedure UPDATEPARAMETERS(w, b, Δw, Δb, count)
2:     w ← w − Δw/count − w · 0.05
3:     b ← b − Δb/count
4:     Δw ← 0, Δb ← 0, count ← 0           ▷ Reset gradient accumulators
5:     return w, b
6: end procedure
```

## 4.2 Evaluation Results

The evaluation was systematically carried out through two sequential experiments, encompassing a diverse array of multidimensional matrix vectors with CKKS.

*4.2.1 Experiment 1: Multidimensional Vector Operations.* This experiment was meticulously designed to perform a comprehensive analysis involving multidimensional matrix vectors:

**Single Dimension-Decimal Points Comparison** In this particular phase, arithmetic operations were executed on single-dimensional matrix vectors of size 10. For each parameter, we choose different values indicated in [22] as a reasonable spectrum in practice. We fix the decimal point of the values in the original vectors (i.e. plain text) and randomly generate floating-point values from (-1, 1) to represent the scientific data that is normalized. We present the results for three vector operations, namely `addition`, `subtraction`, and `multiplication`. To provide a clear estimation of the results, we use the averaged matching decimal points (i.e. across 10 elements in the vector) to indicate how accurate the results are with/without CKKS scheme (note: the integer parts are always identical).

Table 1 shows the summary of the matching decimals of the results for each type of vector operation with CKKS scheme. The rows are ordered based on the matching decimal points. It shows that the cases where there is a larger global scale are likely to lead to a higher number of matching decimal points. In addition, for all three operators, the polynomial degree of 8192 seems to consistently produce the highest number of matching decimal points.

**MultiDimensional Analysis**

Next, we focus on conducting the matrix-matrix multiplication with CKKS scheme. The investigation was expanded to include the comparison of multidimensional vectors in a series of sizes. A thorough collection and analysis of results were conducted, culminating in Figure 1, where we pick the best parameter combination from Table 1. We compute the averaged mean square error and Euclidean distance between the original resulting matrix and the matrix decrypted from the CKKS scheme. Figure 1 shows that two matrices have marginal differences in terms of Euclidean distance and near-to-zero MSE across all sizes of the matrices considered. This evaluation results provide a promising view on applying CKKS to matrix multiplication, one of the dominant building blocks of scientific computation.

these parameters, yielding multiple scenarios of addition, subtraction, and multiplication operations. In the Other experiment, the data consisted of a wildfire dataset. The data used in this research consists of historical wildfires, weather conditions, vegetation indexes, and weather forecasts. These datasets were collected and categorized into different regions and then subjected to preprocessing steps to handle missing values and properly format dates. An example of the datasets includes:

Historical Wildfires: Records of past wildfire occurrences, including the date and region. Weather Data: Information about weather conditions such as temperature, area, min, max, mean, and variance. Vegetation Index: Historical vegetation index information. Weather Forecasts: Forecasts of weather conditions for specific regions and times.

*4.2.2 Experiment 2: Wildfire Dataset.* Informed by the results of Experiment 1, this experiment tests using the CKKS scheme with a wildfire dataset, consisting of historical wildfires, weather conditions, vegetation indexes, and weather forecasts to test its efficacy and to understand how the CKKS encryption scheme could improve security in real-world applications.

**Comparison of Model Accuracy** The comparison of different models, including Linear Regression, Normalized Linear Regression, and Decision Trees, both in their plain and encrypted forms, was performed using the Coefficient of Determination $R^2$ as the metric. The results are shown in Figure 2. The baseline Linear Regression has an $R^2$ value of 0.11, while the encrypted version is slightly lower at 0.10. Normalized Linear Regression exhibits a higher value of 0.31 for the baseline and 0.30 for the encrypted version. The Decision

| ArithmeticOperation | Coefficient Modulus Bit Size | polynomial Modulus Degree | Global Scale | Averaged Matching Decimal Points |
|---|---|---|---|---|
| * | 54 | 2048 | 2**16 | 0.0 |
| * | 75 | 4096 | 2**20 | 0.0 |
| * | 90 | 4096 | 2**25 | 0.0 |
| * | 122 | 8192 | 2**21 | 0.0 |
| * | 206 | 8192 | 2**40 | 0.25 |
| * | 206 | 8192 | 2**21 | 0.375 |
| * | 300 | 32768 | 2**40 | 5.0 |
| * | 200 | 16384 | 2**40 | 6.0 |
| * | 160 | 8192 | 2**40 | 6.5 |
| * | 200 | 8192 | 2**40 | 6.5 |
| + | 54 | 2048 | 2**16 | 0.0 |
| + | 75 | 4096 | 2**20 | 2.0 |
| + | 206 | 8192 | 2**21 | 2.438 |
| + | 122 | 8192 | 2**21 | 3.0 |
| + | 90 | 4096 | 2**25 | 4.0 |
| + | 300 | 32768 | 2**40 | 4.0 |
| + | 206 | 8192 | 2**40 | 4.25 |
| + | 200 | 16384 | 2**40 | 5.0 |
| + | 160 | 8192 | 2**40 | 5.438 |
| - | 122 | 8192 | 2**21 | 1.0 |
| - | 75 | 4096 | 2**20 | 2.0 |
| - | 206 | 8192 | 2**21 | 2.75 |
| - | 90 | 4096 | 2**25 | 4.0 |
| - | 300 | 32768 | 2**40 | 4.0 |
| - | 200 | 16384 | 2**40 | 4.0 |
| - | 160 | 8192 | 2**40 | 4.875 |
| - | 200 | 8192 | 2**40 | 5.167 |
| - | 206 | 8192 | 2**40 | 6.0 |

Table 1: Comparison of Accuracy among different trends of CKKS parameters

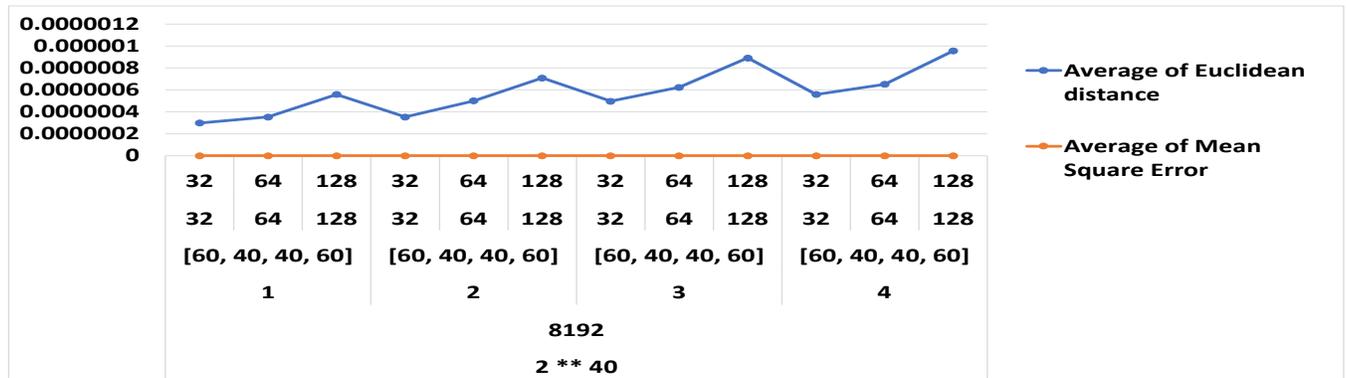

Figure 1: Mean Square Error and Euclidean Distance

Tree has the highest baseline value of 0.40, with the encrypted version slightly lower at 0.38.

These results indicate that employing the CKKS encryption scheme allows for computations on encrypted data without a significant loss in accuracy. Different models show minor variations in $R^2$ when encrypted, but none exhibit a substantial drop, supporting the notion that a variety of models can be used with encrypted data without a major loss in accuracy.

This experiment demonstrates the feasibility of applying CKKS encryption to machine learning models, enabling privacy-preserving computation in sensitive areas like healthcare or finance. The slight decrease in $R^2$ for encrypted models may warrant further investigation but does not overshadow the potential benefits of using encrypted data.

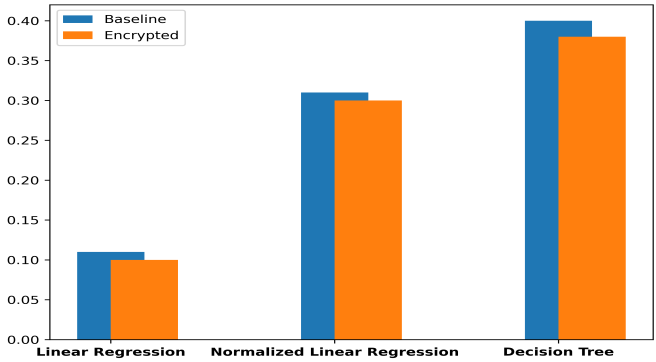

Figure 2: Comparison of Model Accuracy

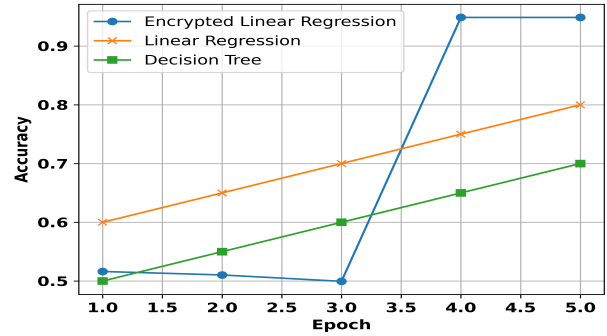

Figure 3: Accuracy Comparison Over Epochs

*4.2.3 Accuracy Comparison Over Epochs.* The accuracy rate over different epochs was analyzed, comparing Encrypted Linear Regression, Linear Regression, and Decision Tree models, as depicted in Figure 3. The accuracy of the Encrypted Linear Regression model starts at around 51.6% and experiences a slight drop before a drastic increase to approximately 94.9% in the fourth epoch, stabilizing thereafter. This behavior may signify the model learning crucial features during the fourth epoch, leading to a substantial accuracy improvement and suggesting convergence.

In contrast, the plain Linear Regression model's accuracy shows a steady increase from 60% to 80% across five epochs, indicating a well-behaved and continuous convergence pattern.

The Decision Tree model also converges steadily but more slowly, with accuracy increasing from 50% to 70%. This more gradual increase may suggest that linear models are better suited for this particular dataset or problem.

**Impacts of Encryption:** Comparing the plain and Encrypted Linear Regression reveals surprising dynamics. The encrypted model's convergence pattern is more erratic, with a sudden jump in accuracy in the fourth epoch, while the plain model converges more steadily. The noise or characteristics introduced by encryption may create these dynamics, but the final higher accuracy of the encrypted model is promising.

The experiment uncovers insights into the convergence behavior of different models in plain and encrypted forms. The erratic behavior of the Encrypted Linear Regression may warrant further investigation, but the result emphasizes the feasibility of using encrypted computation in machine learning. The experiment underscores the potential to achieve competitive performance with encrypted data without significant compromises on accuracy, enhancing security and privacy.

*4.2.4 Overhead Comparison and Optimization Insights.* Detailed insights into the overhead of plain baseline experiments versus encrypted experiments were extracted. The evaluation revealed that while the plain Linear Regression experiment may take less than 10 minutes, encrypted Linear Regression operations and subsequent decryption can require 45 minutes of training.

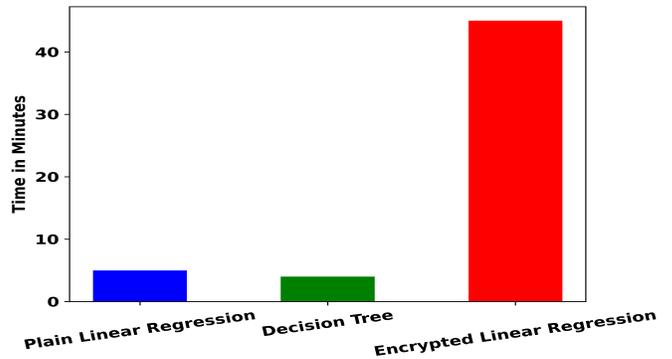

Figure 4: Overhead Comparison and Optimization Insights

We observe that encrypting only the independent variables $X, Y$ significantly reduced the time overhead. The typical time for this experiment was reduced to 45 minutes on our node, as illustrated in Figure 4. The evaluation of encrypted computation both on multidimensional vectors and real-world datasets, such as the wildfire dataset, has been a comprehensive and insightful undertaking. By embracing the complexity inherent in encrypted computations, significant insights were derived, demonstrating the efficacy of this approach in secure data processing. The exploration of the wildfire dataset went beyond theory to emphasize the applicability of the techniques and findings across various scenarios. This not only underlines the robustness of the methodologies but also opens up promising avenues for future research and applications in the field of Super Computing. Also, the use of GPU acceleration could further enhance the findings of this study. By leveraging the parallel processing capabilities of GPUs, the computational time for encrypted models like Tenseal Encrypted Linear Regression can be significantly reduced. This improvement aligns with the real-world need for data privacy without sacrificing efficiency, presenting a promising avenue for advancing encrypted computation paradigms.

## 5 CHALLENGES AND INSIGHTS

*Noise Budget in CKKS.* A significant challenge encountered in our study was the computation of the noise budget in the CKKS (Cheon-Kim-Kim-Song) scheme. While the noise budget is a critical parameter to understand the available room for computations before decryption errors occur, TenSEAL, the library used, does not provide a built-in method to compute it. The lack of this functionality posed limitations on our ability to gauge and control the noise during the experiment, potentially influencing the accuracy of the results.

*Operational Characteristics.* Our findings reveal distinct characteristics for different arithmetic operations. A compromise had to be made to ensure accurate operations at the expense of time and computation cost. The nature of the CKKS scheme made the implementation of division infeasible without compromising encryption.

*Underlying Mechanisms.* In CKKS, real numbers are encoded into polynomials, with the roots representing the input plaintext vector. The encryption of the polynomial into two new polynomials, further incorporating randomized elements, adds more noise. The polynomial operations of addition and multiplication introduce varying degrees of noise, with multiplication being particularly noisy. These inherent characteristics of the CKKS scheme shape the experiment's outcomes and present both challenges and opportunities for optimization.

Future work might focus on developing methods to compute noise budgets in TenSEAL, optimizing the handling of multiplication, and exploring feasible ways to implement division. This exploration contributes to the understanding of the interplay between encryption, arithmetic operations, and the constraints of existing cryptographic tools, illuminating paths for further research and development.

## 6 CONCLUSION

We conduct the extensive examination of encrypted computations through two meticulously designed experiments. We test the practical applications of scientific research within the context of the CKKS scheme. By applying CKKS on various multi-dimensional vector operations, we provide insights to the community on whether CKKS can be useful in the scientific domain, and how much accuracy loss one might expect. Our research sets a valuable first step or ongoing exploration and innovation in Super Computing, aligning with the contemporary emphasis on data privacy and computational efficiency. The insights gained lay a robust foundation for future endeavors, including the exciting possibility of GPU acceleration in CKKS computations within TenSEAL, potentially influencing broader technological advances.